\documentclass{iau} 
\usepackage{graphicx}

\title[ALMA observations of molecular medium in galaxies] 
{High resolution ALMA observations of dense molecular medium in the central regions of active galaxies}

\author[Kotaro Kohno et al.]   
{
Kotaro Kohno$^1$, 
Ryo Ando$^1$, 
Akio Taniguchi$^1$, 
Takuma Izumi$^1$, 
\and 
Tomoka Tosaki$^2$, 
}

\affiliation{
$^1$Institute of Astronomy, The University of Tokyo, Osawa, Mitaka, Tokyo 181-0015, Japan \\ 
email: {\tt kkohno@ioa.s.u-tokyo.ac.jp} \\
$^2$Joetsu University of Education, Yamayashiki, Joetsu, Niigata 943-8512, Japan
}

\pubyear{2015}
\volume{315}  
\jname{From Interstellar Clouds to Star-forming Galaxies: Universal Processes?}
\editors{P.~Jablonka, P.~Andr\'e, \& F.F.S. van der Tak, eds.}
\begin{document}

\maketitle

\begin{abstract} 
In the central regions of active galaxies, dense molecular medium are exposed to various types of radiation and energy injections, such as UV, X-ray, cosmic ray, and shock dissipation. With the rapid progress of chemical models and implementation of new-generation mm/submm interferometry, we are now able to use molecules as powerful diagnostics of the physical and chemical processes in galaxies. Here we give a brief overview on the recent ALMA results to demonstrate how molecules can reveal underlying physical and chemical processes in galaxies. First, new detections of Galactic molecular absorption systems with elevated HCO/H$^{13}$CO$^+$ column density ratios are reported, indicating that these molecular media are irradiated by intense UV fields. Second, we discuss the spatial distributions of various types of shock tracers including HNCO, CH$_3$OH and SiO in NGC 253 and NGC 1068. Lastly, we provide an overview of proposed diagnostic methods of nuclear energy sources using ALMA, with an emphasis on the synergy with sensitive mid-infrared spectroscopy, which will be implemented by JWST and SPICA to disentangle the complex nature of heavily obscured galaxies across the cosmic time.  
\end{abstract}

\firstsection 
\section{ALMA detections of new Galactic molecular absorption systems: HCO as an indicator of intense UV fields}

Molecular medium in front of bright background continuum sources can often be observed as absorption lines, allowing us to make sensitive studies of physical and chemical properties of the interstellar medium (ISM) in various environments, including quiescent and translucent molecular media in the Milky Way, dense molecular gas in the vicinity of active galactic nuclei (AGNs), and the ISM in the early epoch of the universe. In fact, because the absorption line depths are independent of the distance to the absorption system, they offer a very powerful technique to uncover cold molecular gas in the distant universe, which is often difficult to observe in emission unless it is intensely heated by either star-formation or active nuclei. Since the first cosmological detection of such molecular absorption system toward PKS 1413+135 at $z=0.247$ (\cite[Wiklind \& Combes 1994]{Wiklind1994}), it has been demonstrated that molecular absorption systems are extremely useful for studying chemical properties of quiescent ISM (e.g., \cite[Muller \etal\ 2014]{Muller2014}), for determining the redshift of a foreground object in a galaxy-galaxy lens system (\cite[Wiklind \& Combes 1996]{Wiklind1996}), and for constraining fundamental physical quantities and constants such as cosmic background temperature (e.g., \cite[Henkel \etal\ 2009]{Henkel2009}, \cite[Muller \etal\ 2011]{Muller2011}) and proton-electron mass ratios (e.g., \cite[Kanekar \etal\ 2015]{Kanekar2015}) across cosmic time. Despite of their usefulness, however, only four molecular absorption systems beyond $z>0.1$ are currently known at millimeter wavelengths (\cite[Combes 2008]{Combes2008}; \cite[Curran \etal\ 2011]{Curran2011}). 

Another important case is a molecular absorption system in which absorption occurs near an AGN, i.e., presumably in the putative dense obscuring tori around AGNs. For instance, Centaurus A is known to exhibit strong absorption line features seen in both molecules (e.g., CO and HCO$^+$) and neutral hydrogen (e.g., \cite[Espada \etal\ 2010]{Espada2010}; see also Ott \etal\ in this volume). However, the number of such systems, which are believed to originate from such $\sim$pc scale tori, is very small again. NGC 4261 is another example showing nuclear absorption lines in both CO and HI (\cite[Jaffe \& McNamara 1994]{Jaffe1994}) but recent sensitive search for mm-wave absorption features using the Nobeyama Millimeter Array (NMA) and the Plateau de Bure Interferometer (PdBI) yields no significant detections of either CO(1-0) or CO(2-1), giving very small optical depths for these transitions ($\tau_{\rm CO(2-1)} <0.098$, \cite[Okuda \etal\ 2013]{Okuda2013}). 

Such molecular absorption lines are also observed in the diffuse medium within the Galaxy against the background extragalactic continuum sources. This also gives an important way to unveil ``dark'' molecular medium, i.e., invisible in emission because they are often in the equilibrium with the background radiation. 
Currently, the number of such extragalactic continuum sources with known Galactic molecular absorption lines is limited to $\sim$30. Recently, the first detections of HCO, the formyl radical have been reported toward two extragalactic continuum sources (\cite[Liszt \etal\ 2014]{Liszt2014}), opening a new window for studying the impact of UV radiation on these diffuse media because the coexistence of H$_2$ and C$^+$, i.e., the presence of photodissociation regions (PDRs), is required for efficient HCO production (\cite[Gerin \etal\ 2009]{Gerin2009}). Again we need to expand the sample of Galactic absorption systems with such rare species to obtain a comprehensive picture of the diffuse molecular media within the Galaxy, but a survey of new molecular absorption systems is always very time consuming and entails a significant risk of non-detection.

\cite[Ando \etal\ (2015)]{Ando2015} propose using the ALMA archive to search for new molecular absorption systems in mm/submm wavelengths, because the ALMA archive already offers a vast number of sensitive mm/submm spectra of extragalactic continuum sources, i.e., bandpass and/or visibility calibrators, which are essentially radio-loud quasars including BL-LAC objects. Because of the enormous improvement in sensitivity provided by ALMA, these calibrators in the ALMA archive are potentially ideal {\it free lunch} in the quest for molecular absorption systems from the Galaxy to the high-$z$ universe without any new investment of observing time! 

\begin{figure}[b]
\begin{center}
 \includegraphics[width=0.5\textwidth]{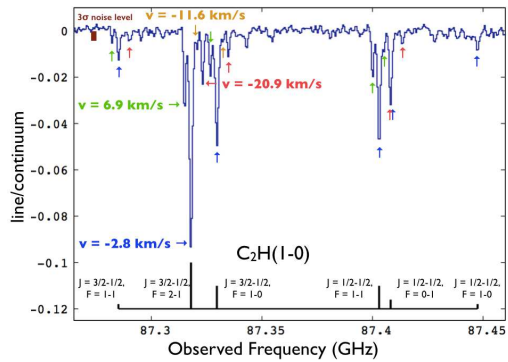} 
 \caption{ALMA band-3 spectrum toward the continuum source J1717-337 showing detections of hyperfine components of C$_2$H in absorption. At least four different velocity components are visible despite of the short integration time (of just a few minutes), demonstrating the superb sensitivity of ALMA. Figure modified from \cite[Ando \etal\ (2015)]{Ando2015}.
 }
   \label{fig1}
\end{center}
\end{figure}

\begin{figure}[b]
\begin{center}
 \includegraphics[width=0.5\textwidth]{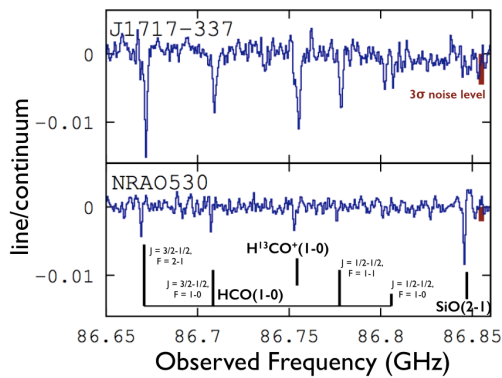} 
 \caption{HCO radical and H$^{13}$CO$^+$ absorption lines toward J1717-337 and NRAO530. Figure modified from \cite[Ando \etal\ (2015)]{Ando2015}.
 }
   \label{fig2}
\end{center}
\end{figure}

\begin{figure}[b]
\begin{center}
 \includegraphics[width=0.5\textwidth]{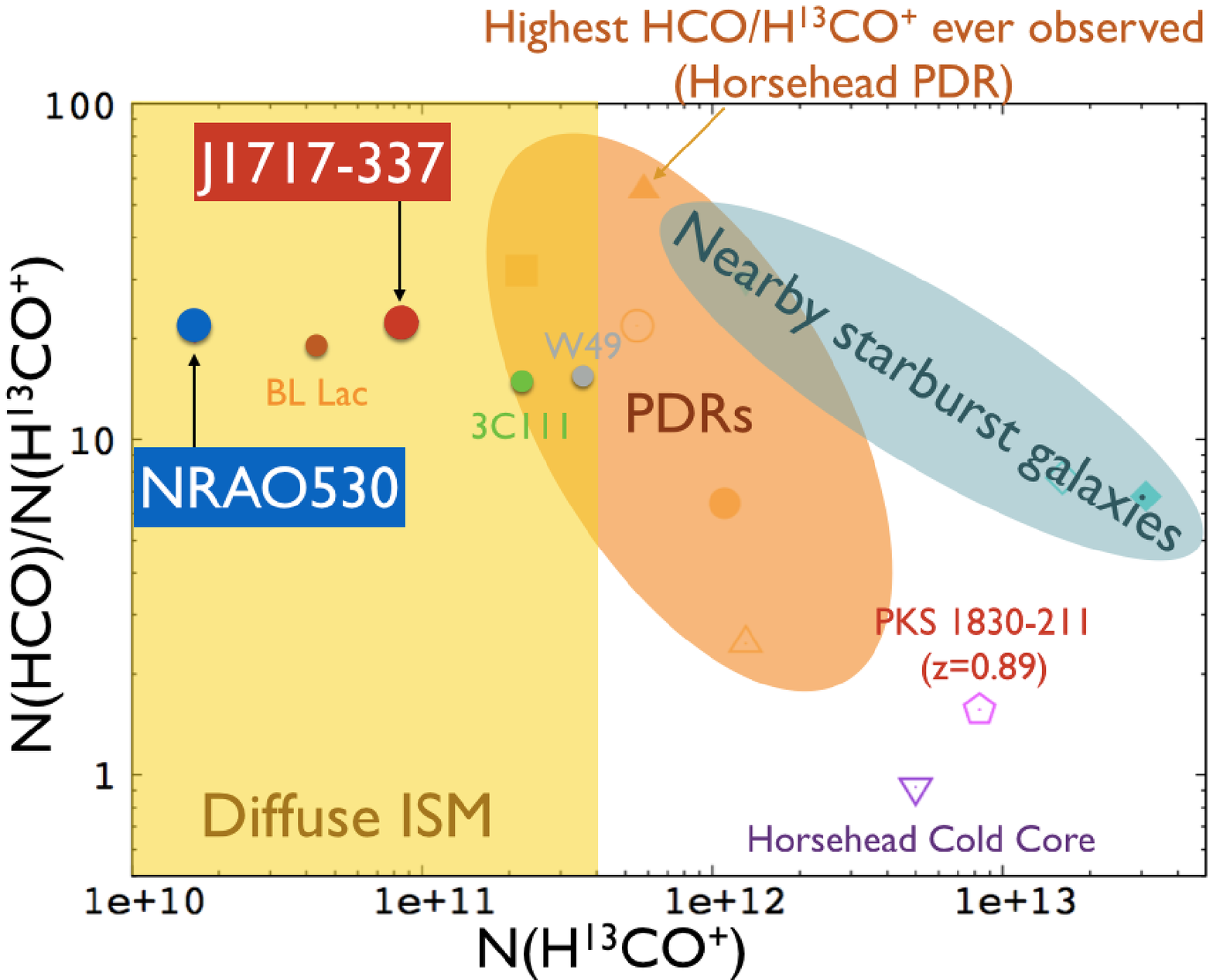} 
 \caption{Column density ratios of HCO to H$^{13}$CO$^+$ as a function of the column density of H$^{13}$CO$^+$, including two new detections of HCO absorption (J1717-337 and NRAO530; \cite[Ando \etal\ 2015]{Ando2015}). Molecular absorption systems toward J1717-337, NRAO530, 3C111, and BL-LAC show elevated HCO/H$^{13}$CO$^+$ column density ratios similar to the value in the Horsehead Nebula (\cite[Gerin \etal\ 2009]{Gerin2009}), indicating that the diffuse media in the Galaxy seen toward these radio sources are exposed in the intense UV field. Figure modified from \cite[Ando \etal\ (2015)]{Ando2015}.
 }
   \label{fig3}
\end{center}
\end{figure}

They analyzed 36 calibrator sources in the ALMA archive to create 3D cubes and search for any absorption features. This resulted in four new detections of absorption line features, and three of which are found to be newly detected Galactic absorption systems. Another one is NRAO530, which is a known absorption system, but strong HCO absorption lines have been detected toward NRAO530 for the first time (Figures \ref{fig1} and \ref{fig2}). With these detections, this doubles the number of Galactic molecular absorption systems exhibiting HCO absorption features, and it was shown that these Galactic diffuse molecular media are in an extreme condition similar to the Horsehead Nebula, a prototypical PDR (Figure \ref{fig3}). 

\section{Shock tracers in the central regions of active galaxies: HNCO, CH$_3$OH, and SiO in NGC 253 and NGC 1068}

SiO is considered to be a tracer of shocks, although it could also be  related to the presence of X-ray radiation (e.g., \cite[Garc\'ia-Burillo \etal\ 2000]{Garcia-Burillo2000}). SiO abundance is enhanced by ejection of significant Si from sputtered silicate grain cores in high velocity shocks (e.g., \cite[Mart\'in-Pintado \etal\ 1992]{Martin-Pintado1992}). In contrast, HNCO forms in the grain mantle, and it can be sublimated in weaker shock events; therefore the HNCO/SiO ratio can be an indicator of shock strength (see also Kelly et al. in this volume). HNCO is also known to be fragile, being easily photodissociated by UV radiation. It is then proposed that the HNCO/CS ratio can give an indicator of starburst evolutionary stage balancing between shock and photodissociation (\cite[Mart\'in \etal\ 2009]{Martin2009}). Methanol, CH$_3$OH, is formed through grain-surface reactions. A series of hydrogenations starting with CO in cold ($\sim$10 K) dust (\cite[Watanabe \& Kouchi 2002]{Watanabe2002}) forms CH$_3$OH efficiently, which then sublimes into the gas phase as the result of shock waves (\cite[Viti \etal\ 2011]{Viti2011}) or warms up dust grains in hot cores (\cite[Garrod \etal\ 2008]{Garrod2008}). It is also suggested that methanol will be easily dissociated by cosmic-ray and UV radiation (e.g., \cite[Aladro \etal\ 2013]{Aladro2013}).

Recent ALMA observations of the central few-100-pc regions of NGC 253 depict the spatial variation of these shock tracers; SiO(2-1) shows a series of clumps with a significant peak at the very center of NGC 253, where an intense H40$\alpha$ recombination line is observed. Meanwhile, HNCO has no significant peak at the very center of the galaxy, suggesting that HNCO is photodissociated by intense UV radiation (\cite[Meier \etal\ 2015]{Meier2015}), although the overall extent of both emissions is quite similar. 

\begin{figure}[b]
\begin{center}
 \includegraphics[width=0.7\textwidth]{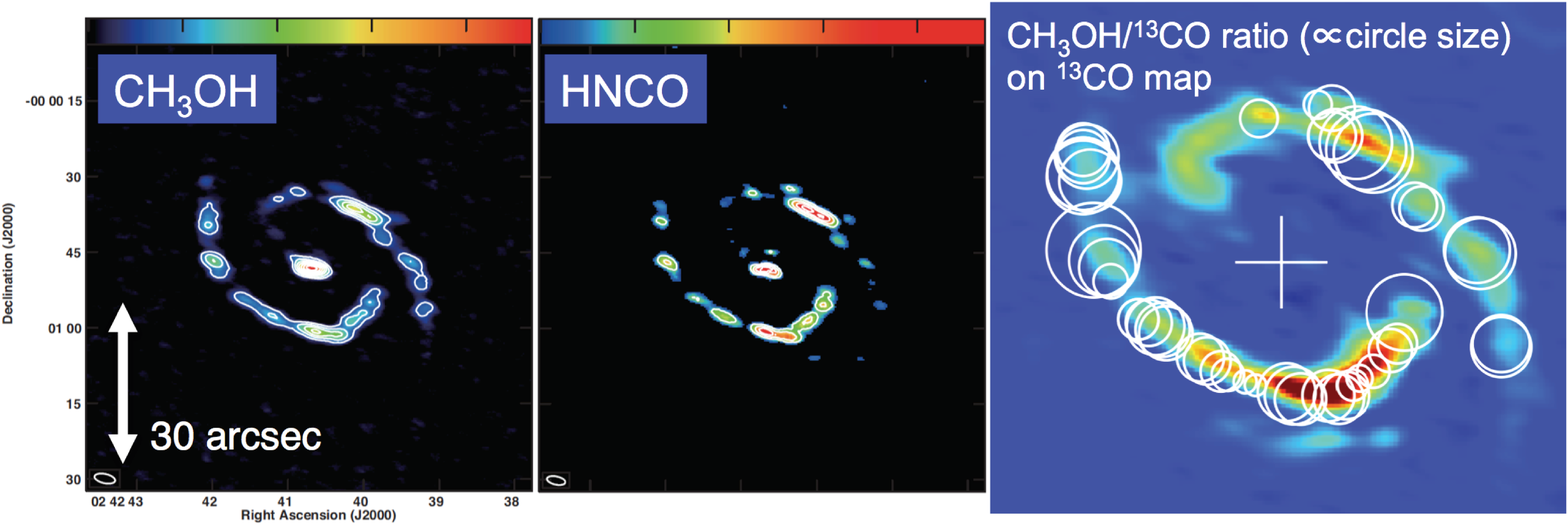} 
 \caption{Distribution of CH$_3$OH (left) and HNCO (middle), tracers of mild shock, in the central $\sim$1 kpc region of NGC 1068. The CH$_3$OH/$^{13}$CO(1-0) intensity ratios, indicated as the sizes of circles superposed on the $^{13}$CO(1-0) integrated intensity image, are displayed in the right panel. Figures adopted from Tosaki \etal\ (in prep.).
 }
   \label{fig4}
\end{center}
\end{figure}

In contrast, in the central $\sim$kpc region of NGC 1068, there are striking differences between tracers of mild and strong shocks. \cite[Takano \etal\ (2014)]{Takano2014} presented the first methanol image of NGC 1068, showing clumps across the starburst ring or arms, as seen in the case of CO and its isotopologs, suggesting that mild shocks are ubiquitous in the arms hosting bursts of massive star formation. Meanwhile, PdBI observations reveal that SiO is concentrated near the circumnuclear disk (CND), without any significant emission across the starbursting spiral arms (\cite[Garc\'ia-Burillo \etal\ 2010]{Garcia-Burillo2010}; see also Kelly, Viti \& Garcia-Burillo in this volume for new SiO(3-2) and HNCO(6-5) data taken with PdBI). New ALMA cycle 2 images of CH$_3$OH and HNCO with improved sensitivity are displayed in Figure \ref{fig4} (Tosaki \etal\ in prep.). We find a striking similarity between CH$_3$OH and HNCO, confirming that mild shocks are ubiquitous in the starbursting spiral arms. It is also evident that these two mild shock tracers are clearly detected in the CND, although the modeling of physical and chemical properties of the CND is not straightforward (\cite[Viti \etal\ 2014]{Viti2014}; \cite[Nakajima \etal\ 2015]{Nakajima2015}). 

A new SiO multi-transition study obtaining four transitions of SiO with ALMA in the central region of NGC 1068 is in progress (Taniguchi \etal\ in prep.), and a preliminary $\sim0''.5$ resolution ALMA image shows that SiO(6-5) is concentrated at the western knot of the CND and no significant emission can be seen in any other regions including the eastern knot, molecular ridges along the inner bar seen in $^{12}$CO (e.g., \cite[Schinnerer \etal\ 2000]{Schinnerer2000}; \cite[Garc\'ia-Burillo \etal\ 2014]{Garcia-Burillo2014}) and spiral arms. This would suggest that the shock properties are drastically different between the Western knot and other regions. Further quantitative analysis of SiO fractional abundance distribution employing rotation diagrams of SiO, CS, and $^{13}$CO are on-going (Taniguchi \etal\ in prep.) to understand the origin of such sharp and anisotropic variation of SiO properties.

\section{Unveiling the nature of obscured galaxies}

With the enormous improvement of the sensitivity and spatial resolution in mm/submm interferometry accomplished by ALMA, it is now strongly expected that ALMA will provide unique diagnostic probes of energy sources in heavily obscured galaxies by penetrating deep dust condensations via mm/submm imaging and/or spectroscopy. 

{\underline{\it IR luminosity density and L/M ratio:}} One of the promising methods is to perform high spatial resolution measurements of the dust continuum to constrain the IR luminosity density and luminosity-to-mass (L/M) ratio in the heart of obscured galaxies. Theoretically, an extreme starburst, which will expel the ISM via radiation pressure onto dust and quench star-formation there, is estimated to have a maximum luminosity surface density of $\sim10^{13}$ $L_\odot$ kpc$^{-2}$ and L/M ratio of $\sim$1,000 $L_\odot/M_\odot$ (\cite[Thompson \etal\ 2005]{Thompson2005}; \cite[Sakamoto \etal\ 2008]{Sakamoto2008}), so spatially resolved measurements of these values would be crucial to test whether such galaxies are powered by a non-stellar source. In fact, PdBI and SMA observations of double nuclei in Arp 220 suggest that the western nucleus of Arp 220 can be heated by an AGN rather than a starburst (\cite[Downes \& Eckart 2007]{Downes2007}; \cite[Sakamoto \etal\ 2008]{Sakamoto2008}). Recent ALMA band-9 observations of the 430$\mu$m dust continuum at a resolution of $0''.36 \times 0''.20$ suggest that the IR luminosity density and the L/M ratio reach $2\times10^{14}$ $L_\odot$ kpc$^{-2}$ and 540 $L_\odot/M_\odot$, respectively, in the western nucleus of Arp 220, indicating the presence of a Compton-thick, heavily obscured AGN at least in the western nucleus of Arp 220 (\cite[Wilson \etal\ 2014]{Wilson2014}).

{\underline{\it the HI/HeII recombination line ratio:}}
Another promising, physically clean indicator would be (the ratio of) recombination lines of H, He, and He$^+$ because they have significantly different ionization potentials (13.6 to 54.4 eV). Specifically, \cite[Scoville \& Murchikova (2013)]{Scoville2013} demonstrate that the H26$\alpha$ (353.623 GHz) to HeII42$\alpha$ (342.894 GHz) luminosity ratio will show a factor of 50 difference between an AGN and a starburst. These two recombination lines can be simultaneously observed with ALMA band-7 with a reasonable integration time for local IR-luminous galaxies, and they will provide an ambiguous basis to calibrate other proposed diagnostic methods. 

{\underline{\it the HCN enhancement:}} A significant enhancement of HCN emission with respect to other molecules such as CO and HCO$^+$ at the centers of local AGNs was first indicated by \cite[Jackson \etal\ (1993)]{Jackson1993} and \cite[Tacconi \etal\ (1994)]{Tacconi1994} based on the NMA and PdBI observations of HCN(1-0) in NGC 1068. 
Since then, mounting supporting evidence for such elevated HCN emission at the centers of local Seyfert galaxies has been accumulated (e.g., \cite[Kohno \etal\ 2001]{Kohno2001}; \cite[Imanishi \etal\ 2006]{Imanishi2006}; \cite[Krips \etal\ 2008]{Krips2008}), giving an empirical AGN diagnostic. There are on-going debates on the origin of such elevated HCN, however. X-ray induced chemistry is one of the attractive scenarios because the CND of NGC 1068 is suggested to be a giant X-ray dominated region (XDR; \cite[Maloney \etal\ 1996]{Maloney1996}) based on detections of HOC$^+$ and other molecules that are preferentially enhanced in the XDR (\cite[Usero \etal\ 2004]{Usero2004}), although chemical models often do not reproduce such HCN enhancement in the XDR (e.g., \cite[Meijerink\etal\ 2007]{Meijerink2007}). 
More recently, spatially resolved ALMA observations of HCN(4-3) and HCO$^+$(4-3) at the CND of NGC 1068 at a $\sim$50 pc resolution reveal that the HCN(4-3)/HCO$^+$(4-3) line ratios are indeed highly elevated at the CND but not significantly at the very position of the active nucleus (\cite[Garc\'ia-Burillo \etal\ 2014]{Garcia-Burillo2014}; \cite[Viti \etal\ 2014]{Viti2014}). This indicates that HCN enhancement is not simply caused by X-ray irradiation. In fact, recent HCN(4-3)/HCO$^+$(4-3) measurements of the high-luminosity type-1 AGN NGC 7469 (\cite[Izumi \etal\ 2015a]{Izumi2015a}) with $L_{\rm X} = 2 \times 10^{43}$ erg s$^{-1}$ suggest that the degree of HCN enhancement is not very different from that in NGC 1097 (\cite[Izumi \etal\ 2013]{Izumi2013}), a much lower luminosity type-1 AGN ($L_{\rm X} =7\times10^{40}$ erg s$^{-1}$), also adding further evidence against the XDR origin of enhanced HCN in AGNs (see also \cite[Mart\'in \etal\ 2015]{Martin2015}). Mechanical heating, which is caused by the dissipation of shock-driven energy (\cite[Loenen \etal\ 2008]{Loenen2008}) and high-temperature-induced chemistry (\cite[Harada \etal\ 2010]{Harada2010}; \cite[Harada \etal\ 2013]{Harada2013}; \cite[Izumi \etal\ 2013]{Izumi2013}) are other possible mechanisms that could explain such HCN enhancement. Elevated HCN/HCO$^+$ ratios in massive (and presumably shocked) molecular outflows driven by the  powerful AGN in Mrk 231 would be consistent with this view (\cite[Aalto \etal\ 2012]{Aalto2012}). A possible HCN maser amplification is also proposed (\cite[Matsushita \etal\ 2015]{Matsushita2015}) in the case of the HCN enhancement in M51 (\cite[Kohno \etal\ 1996]{Kohno1996}). 

\begin{figure}[b]
 \vspace*{-0.3 cm}
\begin{center}
 \includegraphics[width=0.65\textwidth]{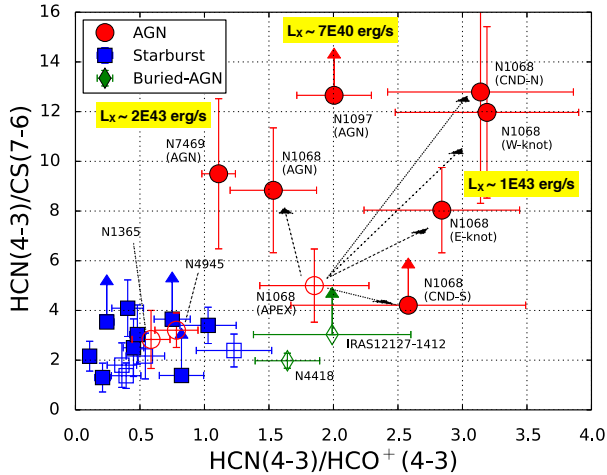} 
 \vspace*{-0.3 cm}
 \caption{HCN(4-3)/HCO$^+$(4-3) and HCN(4-3)/CS(7-6) line ratios of local AGNs, starburst galaxies, and IR-luminous galaxies with buried AGNs. All these lines can be simultaneously observed using ALMA Band-7. In NGC 1068, these line ratios significantly depend on the size of the observing beam; a measurement with APEX ($\sim$1 kpc scale beam; \cite[Zhang \etal\ 2014]{Zhang2014}) is shown as an open red circle, whereas spatially resolved ALMA measurements ($\sim$50 pc; \cite[Garc\'ia-Burillo \etal\ 2014]{Garcia-Burillo2014}) are indicated by red filled circles. It was found that the HCN enhancement at the nucleus of NGC 1068 is not very significant compared with those in the peaks of the CND at $\sim$50 pc resolution, but it is still distinguishable from purely star-forming galaxies, suggesting that these ratios can give an empirical diagnostic of the dominant power sources within the observing beam. Note that NGC 1365 and NGC 4945 (two red open circles), with ratios similar to starbursts, are measured with a single dish telescope (and therefore dominated by circumnuclear starburst). Figure adopted from Izumi \etal\ (2015b).
 }
   \label{fig5}
\end{center}
\end{figure}

Figure \ref{fig5} displays the latest compilation of measurements of HCN(4-3)/HCO$^+$(4-3) and HCN(4-3)/CS(7-6) line ratios currently available, presented by \cite[Izumi \etal\ (2015b)]{Izumi2015b}. This plot demonstrates that the HCN enhancement is still visible even in the $\sim$50 pc resolution ALMA measurements by \cite[Garc\'ia-Burillo \etal\ (2014)]{Garcia-Burillo2014} if we compare those values with star-forming and starburst galaxies. Although the origin of HCN enhancement is yet controversial (see also \cite[Viti 2015, this volume]{Viti 2015}), more (spatially resolved) observational work with ALMA will improve our understanding of such physical and chemical processes and result in identification of useful and reliable tracers of power sources in dusty galaxies.

{\underline{\it Vibrationally excited HCN and synergy with mid-infrared spectroscopy:}} Regarding the nature of HCN enhancement, it is also important to consider the role of intense mid-IR radiation because HCN can be pumped by 14$\mu$m radiation, indicating the indispensable synergy between mm/submm and mid-IR facilities such as JWST and SPICA. Vibrationally excited HCN has been detected in IR-luminous galaxies such as NGC 4418, Mrk 231, and Arp 220 (e.g., \cite[Sakamoto \etal\ 2010]{Sakamoto2010}; \cite[Costagliola \etal\ 2015]{Costagliola2015}; \cite[Aalto \etal\ 2015a,b]{Aalto2015a}). Mid-IR pumping seems to play a significant role in the enhanced high-$J$ HCN emission in the $z=3.91$ quasar APM 08279+5255 (e.g., \cite[Riechers \etal\ 2010]{Riechers2010}). 

\begin{figure}[b]
 \vspace*{-0.3 cm}
\begin{center}
 \includegraphics[width=\textwidth]{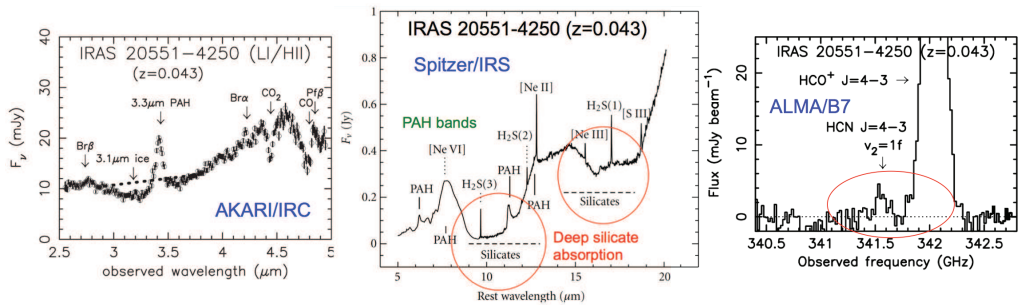} 
 \vspace*{-0.3 cm}
 \caption{Mid-IR and submm spectra of IRAS 20551-4250, a heavily obscured ULIRG at $z=0.043$. Vibrationally-excited HCN emission has been detected (right). Figures taken from \cite[Imanishi \etal\ (2010)]{Imanishi2010}, \cite[Sani \& Nardini (2012)]{Sani2012}, and \cite[Imanishi \& Nakanishi (2013)]{Imanishi2013}.
 }
   \label{fig6}
\end{center}
\end{figure}

Figure \ref{fig6} presents a collection of submm (taken with ALMA) and mid-IR spectra (AKARI and Spitzer) toward the local ultraluminous IR galaxy IRAS 20551-4250, exhibiting vibrationally excited HCN emission (\cite[Imanishi \& Nakanishi 2013]{Imanishi2013}). Because vibrationally excited HCN is preferentially found at heavily obscured nuclei with highly elevated mid-IR luminosity (e.g., \cite[Aalto \etal\ 2015b]{Aalto2015b}), deep silicate absorption features at 9.7 $\mu$m and 18 $\mu$m (indicating $A_{\rm V}>$50 - 100 mag) are often seen in the mid-IR spectra (although their connection is not yet established). Furthermore, PAH band features (the brightest one being at 7.7 $\mu$m) are also useful to constrain the relative contribution of AGN to the total IR luminosity in such dusty nuclei (e.g., \cite[Men\'endez-Delmestre \etal\ 2009]{Menendez-Delmestre2009} and references therein; see also \cite[Privon \etal\ (2015)]{Privon2015} for an example of joint analysis of mm-wave HCN(1-0) and mid-IR properties.). 
JWST will be a powerful tool for studying such mid-IR radiation for local to low-$z$ galaxies, whereas SPICA, equipped with SMI (for 12 - 36 $\mu$m)\footnote{SMI contains 3 spectrometer modules with different resolutions, including the High Resolution Spectrometer, HRS, allowing us to obtain $R$ = 25,000 spectra at 12 - 17 $\mu$m. Note that JWST-MIRI does not have such high spectral resolution spectroscopic capability. See \cite[Sibthorpe \etal\ 2015]{Sibthorpe2015} for details of SMI and the newly defined SPICA mission, a candidate to ESA's 5th medium class mission.} and SAFARI (for 35 - 210 $\mu$m), will be crucial for higher redshift galaxies to catch these redshifted major mid-IR features.

\begin{thebibliography}{}

\bibitem[Aalto \etal\ (2012)]{Aalto2012}
{Aalto, S., Garc\'ia-Burillo, S., Muller, S., et al.} 2012,
\textit{A\&A}, 537, A44

\bibitem[Aalto \etal\ (2015a)]{Aalto2015a}
{Aalto, S., Garc\'ia-Burillo, S., Muller, S., et al.} 2015,
\textit{A\&A}, 574, A85

\bibitem[Aalto \etal\ (2015b)]{Aalto2015b}
{Aalto, S., Costagliola, S., Mart\'in, S., et al.} 2015,
\textit{A\&A}, in press (arXiv:1504.06824) 

\bibitem[Aladro \etal\ (2013)]{Aladro2013}
{Aladro, R., Viti, S., Bayet, E., et al.} 2013,
\textit{A\&A}, 549, A39 

\bibitem[Ando \etal\ (2015)]{Ando2015}
{Ando, R., Kohno, K., Tamura, Y., Izumi, T., et al.} 2015, 
\textit{PASJ}, in press (arXiv:1510.05004) 

\bibitem[Sibthorpe \etal\ (2015)]{Sibthorpe2015}
{Sibthorpe, B., Helmich, F., Roelfsema, P., Kaneda, H., Shibai, H.,  \& the SPICA consortium}, 2015, in 
 \textit{Conditions and Impact of Star Formation From Lab to Space} (EAS Publications Series), in press.

\bibitem[Combes (2008)]{Combes2008}
{Combes, F.} 2008,
\textit{Ap\&SS}, 313, 321

\bibitem[Costagliola \etal\ (2015)]{Costagliola2015}
{Costagliola, F., Sakamoto, K., Muller, S., et al.} 2015, 
\textit{A\&A}, 582, 91

\bibitem[Curran \etal\ (2011)]{Curran2011}
{Curran, S.J., Whiting, M.T., Combes, F., et al.} 2011,
\textit{MNRAS}, 416, 2143 

\bibitem[Downes \& Eckart (2007)]{Downes2007}
{Downes, D., \& Eckart, A.} 2007, 
\textit{A\&A}, 468, L57

\bibitem[Espada \etal\ (2010)]{Espada2010}
{Espada, D., Peck, A.B., Matsushita, S., et al.} 2010, 
\textit{ApJ}, 720, 666


\bibitem[Garc\'ia-Burillo \etal\ (2010)]{Garcia-Burillo2010}
{Garc\'ia-Burillo, S., et al.} 2010,
\textit{A\&A}, 519, A2 

\bibitem[Garc\'ia-Burillo \etal\ (2014)]{Garcia-Burillo2014}
{Garc\'ia-Burillo, S., et al.} 2014,
\textit{A\&A}, 567, A125 

\bibitem[Garrod \etal\ (2008)]{Garrod2008}
{Garrod, R.T., Weaver, S.L.W., \& Herbst, E.} 2008,
\textit{ApJ}, 682, 283

\bibitem[Gerin \etal\ (2009)]{Gerin2009}
{Gerin, M., Goicoechea, J.R., Pety, J., \& Hily-Blant, P.} 2009,
\textit{A\&A}, 494, 977 

\bibitem[Harada \etal\ (2010)]{Harada2010}
{Harada, N., Herbst, E., \& Wakelam, V.} 2010, 
\textit{ApJ}, 721, 1570

\bibitem[Harada \etal\ (2013)]{Harada2013}
{Harada, N., Thompson, T.A., \& Herbst, E.} 2013, 
\textit{ApJ}, 765, 108

\bibitem[Henkel \etal\ (2009)]{Henkel2009}
{Henkel, C., Menten, K.M., Murphy, M.T., et al.} 2009, 
\textit{A\&A}, 500, 725

\bibitem[Imanishi \etal\ (2006)]{Imanishi2006}
{Imanishi, M., Nakanishi, K., \& Kohno, K.} 2006, 
\textit{AJ}, 131, 2888

\bibitem[Imanishi \etal\ (2010)]{Imanishi2010}
{Imanishi, M., Nakagawa, T., Shirahata, M., Ohyama, Y., \& Onaka, T.} 2010, 
\textit{ApJ}, 721, 1233

\bibitem[Imanishi \& Nakanishi (2013)]{Imanishi2013}
{Imanishi, M., \& Nakanishi, K.} 2013, 
\textit{AJ}, 146, 91 

\bibitem[Izumi \etal\ (2013)]{Izumi2013}
{Izumi, T., Kohno, K., Mart\'in, S., et al.} 2013, 
\textit{PASJ}, 65, 100

\bibitem[Izumi \etal\ (2015a)]{Izumi2015a}
{Izumi, T., Kohno, K., Aalto, S., et al.} 2015a, 
\textit{ApJ}, 811, 39

\bibitem[Izumi \etal\ (2015b)]{Izumi2015b}
{Izumi, T., Kohno, K., Aalto, S., et al.} 2015b, 
\textit{ApJ}, in press (arXiv:1512.03438)


\bibitem[Jackson \etal\ (1993)]{Jackson1993}
{Jackson, J.M., Paglione, T.A.D., Ishizuki, S., \& Nguyen-Q-Rieu} 1993, 
\textit{ApJ} (Letters), 418, L13

\bibitem[Jaffe \& McNamara (1994)]{Jaffe1994}
{Croat, T.K., Stadermann, F.J., \& Bernatowicz, T.J.} 2005, 
\textit{ApJ}, 631, 976

\bibitem[Kanekar \etal\ (2015)]{Kanekar2015}
{Kanekar, N., Ubachs, W., Menten, K.M. et al.} 2015, 
\textit{MNRAS}, 448, 104

\bibitem[Kohno \etal\ (1996)]{Kohno1996}
{Kohno, K., Kawabe, R., Tosaki, T., \& Okumura, S.K.} 1996, 
\textit{ApJ} (Letters), 461, L29

\bibitem[Kohno \etal\ (2001)]{Kohno2001}
{Kohno, K., Matsushita, S., Vila-Vilar\'o, B., et al.} 2001, in: J.H. Knapen, J.E. Beckman, I. Shlosman, \& T.J. Mahoney (eds.), 
 \textit{The Central Kiloparsec of Starbursts and AGN: The La Palma Connection} (San Francisco: Astronomical Society of the Pacific), p.\,672
 
 
\bibitem[Liszt \etal\ (2014)]{Liszt2014}
{Liszt, H.S., Pety, J., Gerin, M., \& Lucas, R.} 2014,
\textit{A\&A}, 564, A64


\bibitem[Loenen \etal\ (2008)]{Loenen2008}
{Loenen, A.F., Spaans, M., Baan, W.A., \& Meijerink, R.} 2008,
\textit{A\&A}, 488, L5

\bibitem[Lucas \& Liszt (1998)]{Lucas1998}
{Lucas, R., \& Liszt, H.} 1998, 
\textit{A\&A}, 337, 246

\bibitem[Maloney \etal\ (1996)]{Maloney1996}
{Maloney, P.R., Hollenbach, D.J., \& Tielens, A.G.G.M.} 1996, 
\textit{ApJ}, 466, 561

\bibitem[Mart\'in \etal\ (2009)]{Martin2009}
{Mart\'in, S., Mart\'in-Pintado, \& Mauersberger, R.} 2009, 
\textit{ApJ}, 694, 610

\bibitem[Mart\'in \etal\ (2015)]{Martin2015}
{Mart\'in, S., Kohno, K., Izumi, T., et al.} 2015, 
\textit{A\&A}, 573, 116

\bibitem[Mart\'in-Pintado \etal\ (1992)]{Martin-Pintado1992}
{Mart\'in-Pintado, J., Bachiller, R., \& Fuente, A.} 1992, 
\textit{A\&A}, 254, 315

\bibitem[Matsushita \etal\ (2015)]{Matsushita2015}
{Matsushita, S., Trung, Dinh-V., Boone, F., et al.} 2015, 
\textit{ApJ}, 799, 26

\bibitem[Meier \etal\ (2015)]{Meier2015}
{Meier, D.S., Walter, F., Bolatto, A.D., et al.} 2015, 
\textit{ApJ}, 801, 63

\bibitem[Meijerink \etal\ (2007)]{Meijerink2007}
{Meijerink, R., Spaans, M., \& Israel, F.P.} 2007,
\textit{A\&A}, 461, 793

\bibitem[Men\'endez-Delmestre \etal\ (2009)]{Menendez-Delmestre2009}
{Men\'endez-Delmestre, K., Blain, A.W., Smail, I., et al.} 2009, 
\textit{ApJ}, 699, 667

\bibitem[Nakajima \etal\ (2015)]{Nakajima2015}
{Nakajima, T., Takano, S., Kohno, K., et al.} 2015, 
\textit{PASJ} 67, 8

\bibitem[Okuda \etal\ (2013)]{Okuda2013}
{Okuda, T., Iguchi, S., \& Kohno, K.} 2013, 
\textit{ApJ}, 768, 19

\bibitem[Privon \etal\ (2015)]{Privon2015} 
{Privon, G.C., Herrero-Illana, R., Evans, A.S., et al.} 2015, 
\textit{ApJ}, 814, 39


\bibitem[Riechers \etal\ (2010)]{Riechers2010}
{Riechers, D.A., Wei\ss, A., Walter, F., \& Wagg, J.} 2010, 
\textit{ApJ}, 725, 1032

\bibitem[Sakamoto \etal\ (2008)]{Sakamoto2008}
{Sakamoto, K., Wang, J., Wiedner, M.C., et al.} 2008, 
\textit{ApJ} (Letters), 684, 957

\bibitem[Sakamoto \etal\ (2010)]{Sakamoto2010}
{Sakamoto, K., Aalso, S., Evans, A.S., et al.} 2010, 
\textit{ApJ} (Letters), 725, L228

\bibitem[Sani \& Nardini (2012)]{Sani2012}
{Sani, E., \& Nardini, E.} 2012, 
\textit{Advances in Astronomy}, vol.~2012, id.~783451

\bibitem[Schinnerer \etal\ (2000)]{Schinnerer2000}
{Schinnerer, E., Eckart, A., Tacconi, L.J., \& Genzel, R.} 2000, 
\textit{ApJ}, 533, 85

\bibitem[Scoville \& Murchikova (2013)]{Scoville2013}
{Scoville, N. \& Murchikova, L.} 2013, 
\textit{ApJ}, 779, 75

\bibitem[Tacconi \etal\ (1994)]{Tacconi1994}
{Tacconi, L.J., Genzel, R., Blietz, M., et al.} 1994, 
\textit{ApJ} (Letters), 426, L77

\bibitem[Takano \etal\ (2014)]{Takano2014}
{Takano, S., Nakajima, T., Kohno, K., et al.} 2014, 
\textit{PASJ}, 66, 75

\bibitem[Thompson \etal\ (2005)]{Thompson2005}
{Thompson, T.A., Quataert, E., \& Murray, N.} 2005, 
\textit{ApJ}, 630, 167

\bibitem[Usero \etal\ (2004)]{Usero2004}
{Usero, A., Garc\'ia-Burillo, S., Fuente, A., et al.} 2004, 
\textit{A\&A}, 419, 897

\bibitem[Viti \etal\ (2011)]{Viti2011}
{Viti, S., Jimenez-Serra, I., Yates, J.A., et al.} 2011, 
\textit{ApJ} (Letters), 740, L3

\bibitem[Viti \etal\ (2014)]{Viti2014}
{Viti, S., Garc\'ia-Burillo, S., Fuente, A., et al.} 2014, 
\textit{A\&A}, 570, 28

\bibitem[Watanabe \& Kouchi (2002)]{Watanabe2002}
{Watanabe, N., \& Kouchi, A.} 2002, 
\textit{ApJ} (Letters), 571, L173

\bibitem[Wiklind \& Combes (1994)]{Wiklind1994}
{Wiklind, T., \& Combes, F.} 1994, 
\textit{A\&A}, 286, L9

\bibitem[Wiklind \& Combes (1996)]{Wiklind1996}
{Wiklind, T., \& Combes, F.} 1996, 
\textit{Nature}, 379, 139

\bibitem[Wilson \etal\ (2014)]{Wilson2014}
{Wilson, C.D., Rangwala, N., Glenn, J., et al.} 2014, 
\textit{ApJ} (Letters), 789, L36

\bibitem[Zhang \etal\ (2014)]{Zhang2014}
{Zhang, Z.-Y., Gao, Y., Henkel, C., et al.} 2014, 
\textit{ApJ} (Letters), 784, L31

\end{thebibliography}
\end{document}